\documentclass[prl,showpacs,amsmath,twocolumn]{revtex4}  
\newcommand{\beq}{\begin{equation}}
\newcommand{\enq}{\end{equation}}

\usepackage{graphicx}

\begin{document}
\title{Collective stimulated Brillouin scatter}

\author{Alexander O. Korotkevich$^{1}$, Pavel M. Lushnikov$^{1}$ and Harvey A. Rose$^{2,3}$
}

\affiliation{$^1$ Department of Mathematics and Statistics, University of New Mexico, Albuquerque, NM 87131, USA \\
$^2$ New Mexico Consortium, Los Alamos, New Mexico 87544, USA \\
$^3$Theoretical Division, Los Alamos National Laboratory,
  MS-B213, Los Alamos, New Mexico, 87545
}

\email{har@lanl.gov}

\date{
\today
}

\begin{abstract}
We develop a statistical theory of  stimulated Brillouin backscatter (BSBS)  of a spatially and temporally partially incoherent laser beam for laser fusion relevant plasma. We find
a new collective regime of BSBS which has a much larger threshold than the classical threshold of a coherent beam in long-scale-length laser fusion plasma.
We identify two contributions to BSBS convective instability increment. The first is collective with intensity threshold independent of the laser correlation time and controlled by diffraction.
The second is independent of diffraction, it grows with increase of the correlation time and does not have an intensity threshold.
The  instability threshold is inside the typical parameter region of National Ignition Facility (NIF).
We also find that the bandwidth of  KrF-laser-based fusion systems would be large enough to allow additional suppression of BSBS.
\end{abstract}

\pacs{52.38.-r  52.38.Bv}

\maketitle

Inertial confinement fusion (ICF) experiments require propagation of intense laser light through underdense plasma subject to laser-plasma instabilities which can be deleterious for
achievement of thermonuclear target ignition because they can cause the loss of target symmetry, energy and hot electron production \cite{Lindl2004}.
Among laser-plasma instabilities,  backward stimulated Brillouin backscatter (BSBS) has long  been considered a serious danger because the damping threshold of
BSBS of coherent laser beams is  typically several order of magnitude less then the required laser intensity $\sim  10^{15}\mbox{W}/\mbox{cm}^2$ for ICF. BSBS
may result in laser energy retracing its path to the laser optical
system, possibly damaging laser components \cite{Lindl2004,MeezanEtAlPhysPlasm2010}.

Theory of laser-plasma interaction (LPI) instabilities is well developed for coherent laser beam \cite{Kruer1990}.
However, ICF laser beams are not coherent because temporal and spatial beam smoothing techniques are currently used to produce laser beams with
short enough correlation time, $T_c,$  and lengths to suppress speckle self-focusing.
The laser intensity forms a speckle field - a random in space distribution of intensity with transverse correlation length
$l_c\simeq F\lambda_0$ and longitudinal correlation length (speckle length) $L_{speckle}\simeq 7F^2\lambda_0$, where $F$ is the optic $f/\#$ and $\lambda_0=2\pi/k_0$ is the wavelength
(see e.g. \cite{RosePhysPlasm1995,GarnierPhysPlasm1999}). 
There is a long history of study of amplification in random media (see e.g \cite{vedenov1964,PesmeBerger1994} and references there in).  For small laser beam
  correlation time $T_c$, the spatial instability increment
 is given by a Random Phase Approximation (RPA).
 Beam smoothing for ICF typically has $T_c$ much larger than the for the regime of RPA applicability. There are few examples in which the implications of
  laser beam spatial and temporal incoherence have been analyzed for such larger $T_c$. One exception is forward stimulated Brillouin scattering (FSBS). Although FSBS for a strictly coherent
  laser beam is a classic linear theory, we have  obtained \cite{LushnikovRosePRL2004,LushnikovRosePlasmPhysContrFusion2006} its dispersion relation for laser beam
  correlation time $T_c$ too large for RPA relevance, but $T_c$ small enough to suppress single laser speckle instabilities \cite{RoseDuBois1994}.
  We verified our theory of this "collective" FSBS regime with 3D simulations. Similar simulation results had been previously observed \cite{SchmittAfeyan1998}.
  This naturally leads one to consider the possibility of a collective regime for BSBS backscatter (CBSBS). We will present 2D and 3D simulation results as evidence for such a regime, and find
   agreement with a simple theory  that above CBSBS threshold, the spatial increment for
  backscatter amplitude $\kappa_i$, is well approximated by the sum of two contributions.   The first
is RPA-like  $\propto T_c$ without intensity threshold (we neglect light wave damping). The second
 has a threshold as a function of laser intensity.
  For National Ignition Facility NIF parameters
  the  threshold is comparable with  NIF intensities. That second contribution is collective-like because
it neglects speckle contributions and is only weakly dependent on $T_c$.
CSBSB threshold is applicable for strong and weak acoustic damping coefficient $\nu_{ia}$.
The theory also provides a good quantitative prediction of the instability increment for small $\nu_{ia}\sim 0.01$
  which is relevant for gold
plasma near the wall of hohlraum in  NIF experiments\cite{Lindl2004}.

Assume that laser beam propagates in plasma with frequency
$\omega_0$  along $z$. The  electric field
$\cal E$ is given by
\begin{eqnarray}\label{EBdef}
{\cal E}=(1/2)e^{-i\omega_0 t}\Big [E e^{ik_0 z}+Be^{-ik_0
z-i\Delta\omega t}\Big ]+c.c.,
\end{eqnarray}
where $E({\bf r}, z,t)$ is the envelope of laser beam and $B({\bf
r}, z,t)$ is the envelope of backscattered wave,  ${\bf r}=(x,y)$,
and c.c. means complex conjugated terms. Frequency shift $\Delta
\omega=-2k_0c_s$ is determined by coupling of $E$ and $B$ through ion-acoustic wave with phase speed
$c_s$ and wavevector $2k_0$ with  plasma density fluctuation
$\delta n_e$ given by $\frac{\delta n_e}{n_e}=\frac{1}{2}\sigma
e^{2ik_0z+i\Delta\omega t}+c.c.,$ where $\sigma({\bf r}, z,t)$ is
the slow envelope and $n_e$ is the average electron density, assumed
small compared to critical density, $n_c$. The coupling of $E$
and $B$ to plasma density fluctuations gives
\begin{eqnarray}\label{EBeq1}
R_{EE}^{-1}E \equiv  \left [ i\Big (c^{-1}{\partial_t}+{\partial_z}\Big )+\frac{1}{2k_0}\nabla^2
  \right ]E=\frac{k_0}{4}\frac{n_e}{n_c}\sigma B, \\
R_{BB}^{-1}B  
\equiv  \left [ i\Big (c^{-1}{\partial_t}-{\partial_z}\Big )+\frac{1}{2k_0}\nabla^2
  \right ]B=\frac{k_0}{4}\frac{n_e}{n_c}\sigma^* E, \label{EBeq2}
\end{eqnarray}
 $\nabla=({\partial_x},{ \partial_y})$, and $\sigma$ is described by the acoustic wave equation coupled to
the pondermotive force $\propto {\cal E}^2$ which results in the
envelope equation
\begin{eqnarray}\label{sigma1}
R_{\sigma\sigma}^{-1}\sigma^* \equiv   [ i ({c_s^{-1}}{\partial_t}+2\nu_{ia} k_0+{\partial_z} )-(4k_0)^{-1}\nabla^2
   ]\sigma^* \nonumber \\
   =-2k_0  E^*B.
\end{eqnarray}
The response of the slowly varying part of $\delta n_e$ to the slowly varying part of the ponderomotive force, proportional to $|E|^2 + |B|^2$, responsible for self-focusing, is neglected.
$\nu_{ia}=\nu_L/2k_0c_s$ is the scaled acoustic Landau damping coefficient.  $E$ and $B$ are in thermal units (see e.g. \cite{LushnikovRosePRL2004}).

Assume that laser beam was made partially incoherent through induced spacial incoherence beam smoothing \cite{LehmbergObenschain1983} which defines stochastic boundary conditions at $z=0$ for the
spacial Fourier transform (over ${\bf r}$) components $ \hat E({\bf k})$, of laser beam amplitude
\cite{LushnikovRosePRL2004}:
\begin{eqnarray}\label{phik}
\hat E({\bf k },z=0,t)= |E_{\bf k}|\exp [ i\phi_{\bf
k}(t) ], \nonumber \\
  \langle \exp i [\phi_{\bf
k}(t)-\phi_{{\bf k}'}(t') ] \rangle =\delta_{{\bf k
k}'}\exp (-|t-t'|/T_c),
 \nonumber \\
 |E_{\bf k}|=const, \ k<k_m; \  E_{\bf k}=0, \ k>k_m,
\end{eqnarray}
chosen as the idealized "top hat" model of NIF optics \cite{polarization}. Here $k_m\simeq k_0/(2F)$ and the average intensity, $   \langle I
\rangle \equiv  \langle |E|^2 \rangle =I$ determines the constant.

In linear approximation, assuming $|B|\ll |E|$ so that only the laser beam is BSBS unstable, we can neglect  right hand side (r.h.s.) of Eq. (\ref{EBeq1}).
 The
resulting linear equation with top hat boundary condition (\ref{phik}) has the exact solution as
 decomposition of $E$ into
 Fourier series,
$E({\bf r},z,t)=\sum_j E_{{\bf k}_j}$
%
with $E_{{\bf k}_j} \propto \exp\big [ i(\phi_{{\bf k}_j}(t-z/c)+{\bf k}_j\cdot {\bf r}-{\bf k}_j^2z/2k_0)\big ].$

\begin{figure}
\begin{center}
\includegraphics[width = 3.3 in]{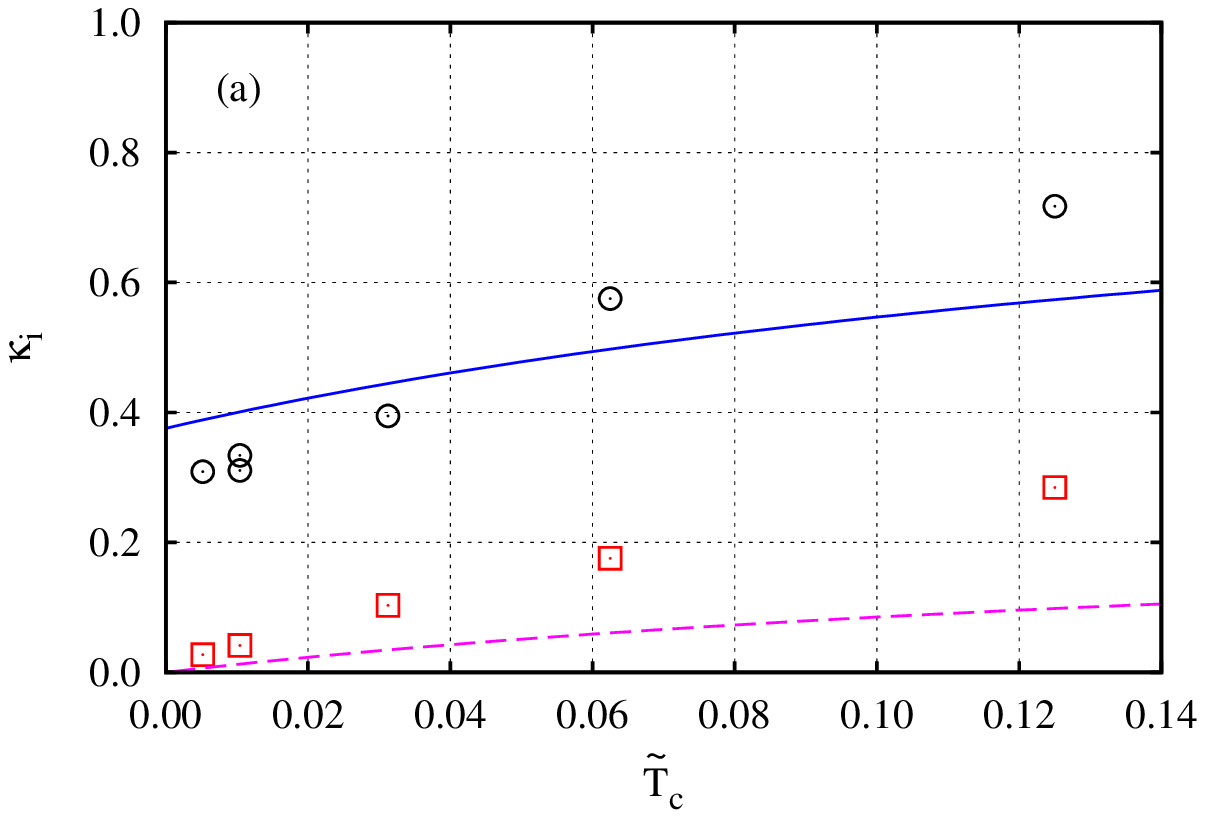}\\
\includegraphics[width = 3.3 in]{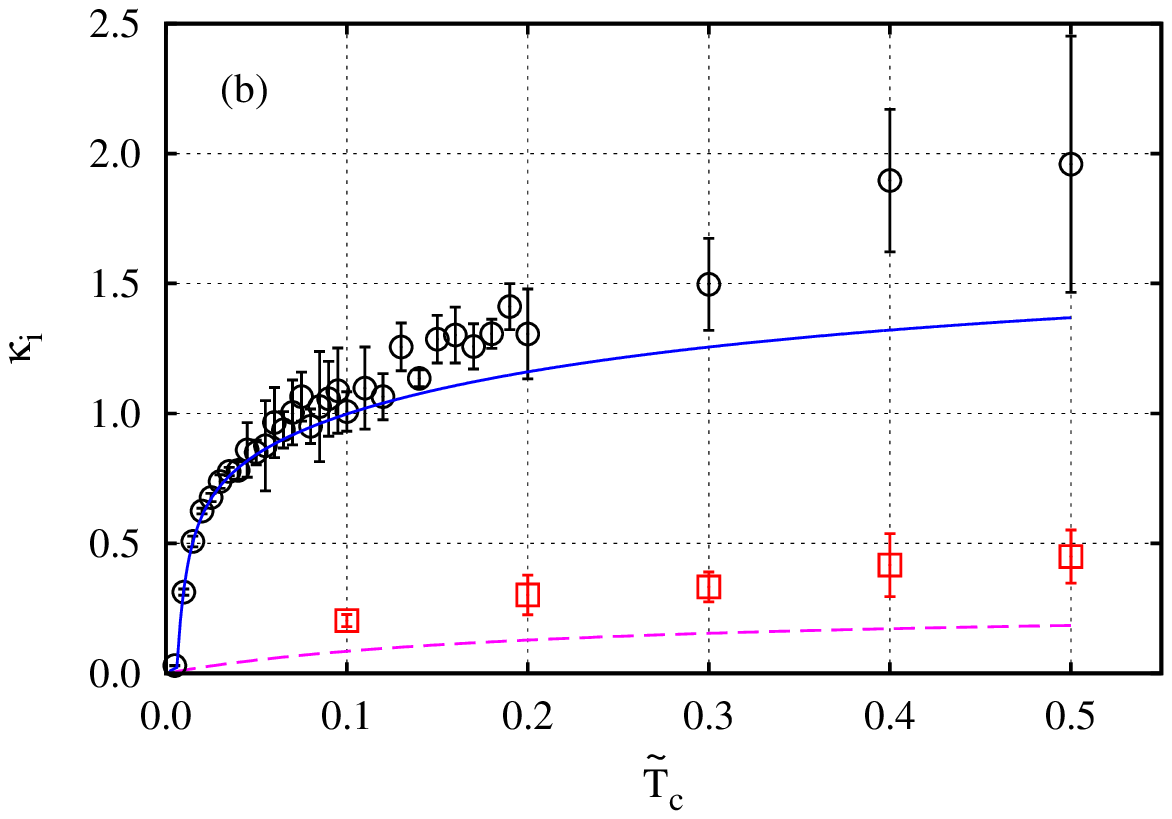}\\
\end{center}
\caption{Spatial increment $\kappa_i$ of CBSBS obtained from numerical simulations compared with the sum of increments
$\kappa_B+\kappa_\sigma$ (obtained by solving (\ref{dispk0contTc}) and (\ref{dispk0contTcB})). Parameters of simulations are $\nu_{ia}=0.01, \ F=8$. (a) $3+1D$ simulations with
$c_s/c=0$, $\tilde I=2$ (circles) and $\tilde I=1$ (squares). Solid and dashed lines show $\kappa_B+\kappa_\sigma$ for  $\tilde I=2$ and $\tilde I=1$, respectively. If $\kappa_\sigma<0$ then
$\kappa_B+\kappa_\sigma$ is replaced by $\kappa_B$.
(b) $2+1D$ simulations   with the modified boundary conditions, $c_s/c=1/500$,   $\tilde I=3$ (circles) and $\tilde I=1$ (squares). Error bars  are also shown.
Solid and dashed lines show $\kappa_B+\kappa_\sigma$ for  $\tilde I=3$ and $\tilde I=1$, respectively.
}
\label{fig:figkappa}
\end{figure}

Figures  \ref{fig:figkappa} show the  increment $\kappa_i$
of the spatial growth of backscattered light intensity $\langle |B|^2\rangle \propto e^{-2\kappa_i z}$ as a
 function of the rescaled correlation time $\tilde T_c\equiv T_c k_0c_s /4F^2$ (note that definition is different by a factor $1/2F$ from the definition used for FSBS
 \cite{LushnikovRosePRL2004,LushnikovRosePlasmPhysContrFusion2006}) obtained from the numerical solution of the linearized equations (\ref{EBeq1})-(\ref{sigma1})
 using operator splitting
method along the characteristics of  $E$ and $B$. Here and below we use dimensionless units with  $k_0/k_m^2$ as the unit in $z$ direction,
$k_0/k_m^2 c_s$ is the time unit and $\mu\equiv 2\nu_{ia} k_0^2/k_m^2,$
Also $\langle \ldots \rangle$ means averaging over the statistics of laser beam fluctuations  (\ref{phik}) and $\tilde I$ is the scaled dimensionless laser intensity defined as
 $\tilde I=\frac{4F^2}{\nu_{ia}}\frac{n_e}{n_c} I$.
 Figure  \ref{fig:figkappa}a corresponds to the $3+1D$ simulations (three spatial coordinates and $t$)  with the boundary and initial conditions (\ref{phik}) in the limit $c\to \infty$
 (i.e., setting $c^{-1}$ terms in (\ref{EBeq1})-(\ref{sigma1}) to be zero. Figure  \ref{fig:figkappa}b
  shows the result of $2+1D$ simulations (only 1 transverse spatial variable is taken into account) for the modified boundary condition
  compare with the last line in (\ref{phik}) as $ |E_{\bf k}|=k^{1/2} \, const, \ k<k_m; \  E_{\bf k}=0, \ k>k_m$ which is chosen to mimic the extra factor $k$ in the
  integral over transverse direction of the full  $3+1D$ problem.   In that case  $c/c_s \simeq 500.$
E.g. for $\tilde T_c=0.1$ we typically use 256 transverse Fourier modes and  a discrete steps $\Delta z = 0.15$ in dimensionless units 
with the
total length of the system $L_z = 100$ and a time step $\Delta t = \Delta z/c$. For each simulation we typically have to wait $\sim 10^6$ time steps to achieve a statistical steady state and then average over next
$\sim 10^6$ time steps to find $\kappa_i$.

We now relate $\kappa_i$ to the instability increments for   $\langle B \rangle$ and $\langle \sigma^* \rangle$ (we designate them $\kappa_B$ and $\kappa_\sigma$, respectively).
In general, growth rates of mean amplitudes only give a lower bound to $\kappa_i$.
First we look for $\kappa_\sigma$.  Eq. (\ref{EBeq2}) is linear in $B$ and $E$  which implies that
$B$ can be decomposed into $B=\sum_j B_{{\bf k}_j}.$
We approximate r.h.s. of (\ref{sigma1}) as $E^*B\simeq \sum_j
E_{{\bf k}_j}^*
B_{{\bf k}_j}$ so that %
\begin{eqnarray}\label{sigma2}
 R_{\sigma\sigma}^{-1}\sigma^*
  =-2k_0  \sum_jE_{{\bf k}_j}^*B_{{\bf k}_j},
\end{eqnarray}
which means that we neglect off-diagonal terms $E_{{\bf k}_j}^*
B_{{{\bf k}_j}'}, \quad j\neq j'.$ Since speckles of
laser field arise from interference of different Fourier modes,
$j\neq j',$ we associate the off-diagonal terms with speckle
contribution to BSBS
\cite{RoseDuBois1993,RosePhysPlasm1995,RoseMounaixPhysPlasm2011}.
The neglect of  off-diagonal terms requires that during time $T_c$ light travels much further than a speckle length, $L_{speckle}\ll cT_c$
  and that $T_c \ll t_{sat}$, where $t_{sat}$ is the characteristic time scale at which
BSBS convective gain saturates at each speckle \cite{MounaixPRL2000}.

Eqs. (\ref{EBeq2}) and (\ref{sigma2}) result in the closed expression
$R_{\sigma\sigma}^{-1}\langle \sigma^*\rangle =-(k_0^2/2)(n_e/n_c)\langle E^*   R_{BB}\sigma^* E  \rangle$ which has the same form as the Bourret
approximation \cite{PesmeBerger1994}. We look for the solution of that expression in exponential form
$B_j, \sigma^* \propto e^{i(\kappa z+{\bf k}\cdot {\bf r}-\omega t)}$, then the exponential time dependence
in (\ref{phik}) allows to carry integrations in that expression explicitly to arrive at the following
relation in dimensionless units
\begin{eqnarray}\label{dispk1}
-i\omega+\mu+i\kappa-(i/4)k^2 \qquad  \qquad  \qquad  \qquad  \qquad  \qquad   \nonumber \\
=8iF^4\frac{n_e}{n_c}\sum\limits_{j=1}^N \frac{|E_{{\bf k}_j}|^2}{\omega\frac{c_s}{c}+\kappa-k_j^2-\frac{k^2}{2}-{\bf k}_j\cdot {\bf k}+2i\frac{c_s}{c}\frac{1}{\tilde T_c}},
\end{eqnarray}
where  $1/k_m$ is the transverse unit of length and vectors ${\bf k}_j$ span the entire top hat (\ref{phik}), i.e. $I=\sum_j|E_{{\bf k}_j}|^2$.

In the continuous limit $N\to \infty$, sum in (\ref{dispk1}) is replaced by integral which gives for the most unstable mode ${\bf k}=0$:
\begin{eqnarray}\label{dispk0contTc}
-i\omega+\mu+i\kappa
+i\frac{\mu}{4}\tilde I\ln\frac{1-\kappa-\omega\frac{c_s}{c}-2i\frac{c_s}{c}\frac{1}{\tilde T_c}}{-\kappa-\omega\frac{c_s}{c}-2i\frac{c_s}{c}\frac{1}{\tilde T_c}}
=0.
\end{eqnarray}
The relation (\ref{dispk0contTc}) supports the convective instability with the increment $\kappa_\sigma\equiv Im(\kappa)>0$ only for $\tilde I> \tilde I_{convthresh}$, where $\tilde I_{convthresh}$
is the convective CBSBS threshold given by
\begin{eqnarray}\label{I0convthresh}
\tilde I_{convthresh}=\frac{4F^2}{\nu_{ia}}\frac{n_e}{n_c}I_{convthresh}=4/\pi.
\end{eqnarray}
In the limit $c/c_s \to \infty$,  the increment $\kappa_\sigma$ is independent of $\tilde T_c$  which suggests that we refer to it as the collective instability branch.
For finite but small $c_s/c\ll 1$ and $\tilde I>\tilde I_{convthresh}$  there is sharp transition of $\kappa_\sigma$ as a function of $\tilde T_c$ from 0 for $\tilde T_c=0$ to
 $\tilde T_c$-independent value of $\kappa_\sigma$. That value can be obtained analytically from (\ref{dispk0contTc})
 for $I$ just above the threshold  as follows: $\kappa_i=\mu(\pi/4)(\tilde I-I_{convthresh})/(\mu \tilde I-1)$.

The increment $\kappa_B$ is obtained in a similar way by statistical averaging of equation (\ref{EBeq2}) for $\langle B\rangle $ with $\sigma^*$  from equation (\ref{sigma1}) which gives
\begin{eqnarray}\label{dispk0contTcB}
-i\omega\frac{c_s}{c}+i\kappa 
+
i\frac{\mu}{4}\tilde I\frac{1}{\kappa-\omega-i\mu-i\frac{1}{\tilde T_c}}
=0.
\end{eqnarray}
Here we neglected the contribution to $\kappa_B\equiv Im(\kappa)$ from diffraction which gives negligible correction.
Equation (\ref{dispk0contTcB}) does not have a convective threshold (provided we neglect here light wave damping) while $\kappa_B$
has near-linear dependence on   $\tilde T_c:$ $\kappa_B\simeq \mu\tilde I\tilde T_c/4$ for $\tilde T_c< 1/\mu$ which is typical for RPA results. It suggests that we refer $\kappa_B$ as   the RPA-like branch of
instability.

We choose $\omega=0.5$ in  (\ref{dispk0contTc}) and $\omega=0$ in (\ref{dispk0contTcB}) to maximize $\kappa_\sigma$ and $\kappa_B$, respectively.
Equation (\ref{dispk0contTc}) 
also predicts absolute instability for $\tilde I> \mu+3\mu^{-1}+O(\mu^{-3})+O(\tilde T_c^{-1} c_s/c)$, which is slightly above the coherent absolute threshold $\tilde I=\mu$ but
here we emphasize the convective regime. 
Figures \ref{fig:figkappa}a and b show  that the analytical expression $\kappa_B+\kappa_\sigma$ is a reasonable good approximation for numerical value of $\kappa_i$ above the convective threshold (\ref{I0convthresh})
for $\tilde T_c\lesssim 0.1$ which is the main result of this Letter. Below this threshold the analytical and numerical results are in qualitative agreement at best but in that case we replace
$\kappa_B+\kappa_\sigma$ by $\kappa_B$ because $\kappa_\sigma<0$ in that case.

The qualitative explanation why $\kappa_B+\kappa_\sigma$ is a surprisingly good approximation to $\kappa_i$ is based on the following argument.
First imagine that $B$ propagates linearly and not coupled to the fluctuations of $\sigma^*$, so its
source is  $\sigma^* E\to \langle \sigma^* \rangle  E$ in r.h.s of (\ref{EBeq2}). 
If
$ \langle \sigma^*  \rangle \propto e^{\kappa_\sigma z}$ grows slowly with $z$ (i.e. if $ \langle \sigma^*  \rangle$ changes a little over the speckle length $L_{speckle}$ and time $T_c$),
then so will  $\langle |B|^2 \rangle$ at the rate $2\kappa_\sigma$ . But if the
total linear response $R_{BB}^{tot}$ ($R_{BB}^{tot}$ is the renormalization of bare response $R_{BB}$ due to the coupling in r.h.s of  (\ref{EBeq2})) is unstable  
then its growth
rate gets added to $\kappa_\sigma$ in the determination of $\langle |B|^2 \rangle$ since in all theories which allow
factorization of 4-point function into product of 2-point functions,
$\langle B(1)B^*(2)\rangle = R^{tot}_{BB} (1,1')S(1',2')R^{tot \, *}_{BB} (2',2)$. Here
$S(1,2) \equiv \langle\sigma^* (1)\sigma (2)\rangle \langle E(1)E^*(2) \rangle  \simeq \langle\sigma^*  (1)\rangle\langle\sigma (2)\rangle \langle E(1)E^*(2) \rangle$
and $"1", \, "2"$ etc. mean 
of all spatial and temporal arguments.

The applicability conditions of the Bourret approximation used in derivation of (\ref{dispk0contTc}) and (\ref{dispk0contTcB}) in the dimensionless units are
\begin{equation}\label{deltaomegadeltaomega}
\Delta \omega_B\Delta \omega_\sigma \gg\gamma_0^2.
\end{equation}
and $\Delta \omega_B\gg (c/c_s)|\kappa_B|$ as well as $\Delta \omega_\sigma\gg\mu$.
Here $\gamma_0$ is the temporal growth rate of the spatially homogeneous solution which is given by $\gamma_0^2=(1/4)(c/c_s)\mu\tilde I.$  Also
$\Delta \omega_\sigma=1/\tilde T_c$ is the bandwidth for $\sigma$  and  $\Delta \omega_B$ is the effective bandwidth for $B$.  $\Delta \omega_B$ is dominated by the diffraction in (\ref{EBeq2})
which gives in the dimensionless units $\Delta \omega_B=c/c_s$.   Then (\ref{deltaomegadeltaomega}) reduces to  $\tilde T_c\ll 4/(\mu \tilde I)$ and $|\kappa_B|\ll 1$.
Together with the condition $T_c\gg L_{speckle}/c$ used in the derivation of (\ref{dispk0contTc}) and assuming that $\tilde I\simeq \tilde I_{convthresh}$, it gives a double inequality $(7\pi/2)(c_s/c)\ll \tilde T_c\ll \pi/\mu$
which can be well satisfied for $\mu\simeq 5$, i.e. for $\nu_{ia}\simeq 0.01$ as in
gold NIF plasma but not for $\mu\simeq 50$ as in low ionization number $Z$ NIF plasma. Also $|\kappa_B|<1$ implies that $\tilde I >\tilde I_{convthresh}$  because otherwise,  below that threshold, $\kappa_B\sim -\mu$ which would
contradict $|\kappa_B|<1$.
All these conditions are  satisfied for $\tilde T_c\lesssim 1/4$ for the parameters of Figure \ref{fig:figkappa} with $\tilde I=2$ or $\tilde I=3$ (solid lines in Figure \ref{fig:figkappa}) but not for $\tilde I=1$
(dashed lines in Figure \ref{fig:figkappa}). Additionally, an estimate for $T_c\ll t_{sat}$ from the linear part of the theory of Ref. \cite{MounaixPRL2000} results in the condition
$\tilde T_c\ll 8\tilde I/\mu$ which is much less restrictive than the previous condition.
  These estimates are consistent with the observed agreement between  $\kappa_i=\kappa_\sigma+\kappa_B$ and $\kappa_i$ from simulations
(filled circles in Figure \ref{fig:figkappa}) for $\tilde I$ above the threshold (\ref{I0convthresh}).
We conclude from Figure \ref{fig:figkappa} that the applicability condition for  the Bourret approximation is close to the domain of $\tilde T_c$ values for which $\kappa_i=\kappa_\sigma+\kappa_B$.

\begin{figure}
\begin{center}
\includegraphics[width = 1.9 in]{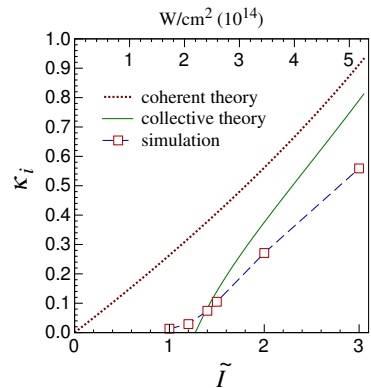}
\end{center}
\caption{$\kappa_i$ vs. $\tilde I$ for $\mu=5.12$  obtained from simulations  (squares connected by dashed line,  $c_s/c=0$ and limit $\tilde T_c\to 0$ taken by extrapolation from $\tilde T_c\ll 1$),
analytical result $\kappa_\sigma$ (solid curve) and coherent laser beam increment  $\kappa_{coherent}$ (dotted curve).
Upper grid corresponds to laser intensity in dimensional units for  NIF parameters and gold plasma
$T_e\simeq 5\mbox{keV}, \ F=8,\  \ n_e/n_c=0.1, \nu_{ia}=0.01, \  \lambda_0=351 \mbox{nm}$.}
\label{fig:fig1}
\end{figure}

 For
typical NIF parameters  \cite{Lindl2004,LushnikovRosePlasmPhysContrFusion2006}, $ F=8,\  \ n_e/n_c=0.1, \ \lambda_0=351 \mbox{nm}$ and $c_s=6\times 10^{7}\ \mbox{cm s}^{-1}$ and the
electron plasma temperature $T_e\simeq 5\mbox{keV}$,
we obtain from (\ref{I0convthresh}) that
$I_{convthresh}\simeq 2.2\times 10^{14}\mbox{W}/\mbox{cm}^2$ for gold plasma with  with $\nu_{ia}\simeq 0.01$, in the range of NIF single polarization intensities.   So we conclude that for gold NIF
plasma $I\sim I_{convthresh}$ while for  low $Z$ plasma with $\nu_{ia}\sim 0.1$ $I$ is well below  $I_{convthresh}$.  Fig. \ref{fig:fig1} shows $\kappa_i$ in the limit $c_s/c=0, \ \tilde T_c\to 0$ from simulations,
analytical result $\kappa_\sigma$ ($\kappa_B=0$ in that limit) and the instability increment of the coherent laser beam $\kappa_{coherent}=\mu/2-(\mu^2-\mu\tilde I)^{1/2}/2$ (see e.g. \cite{Kruer1990}).
It is seen that the coherent increment significantly overestimates numerical increment especially around $I_{convthresh}.$
If we include the effect of finite $c_s/c=1/500$ and finite $\tilde T_c$ as in Fig. \ref{fig:figkappa}b then $\kappa_i$ has a significant dependence on $\tilde T_c$.
Current NIF 3{\AA} beam smoothing design has $T_c\simeq 4$ps which implies $\tilde T\simeq 0.15$. In that case Fig. \ref{fig:figkappa}b
shows  that there is a significant (about 5 fold) change in $\kappa_i$ between $\tilde I=1$ and $\tilde I=3$.
Similar estimate for KrF lasers ($\lambda_0=248 \mbox{nm}, \ F=8, \ T_c=0.7$pm) gives $\tilde T_c =0.04\ \mbox{[ps]}$  which results in a significant ($40\%$) reduction of $\kappa_i$ for $\tilde I=3$
compare with above NIF estimate.

For practical application the threshold of BSBS is often understood as the total gain required to  amplify initial thermal fluctuations up to $|B|^2\sim |E|^2$.
With such definition of threshold our results indicate that the coherent BSBS increment significantly overestimates $\kappa_i$ for practical values of $\tilde T_c$ as can be seen from a comparison of Figures
\ref{fig:figkappa}b and \ref{fig:fig1}.

$T_c$ in NIF can be further reduced  by self-induced temporal incoherence 
 with  collective FSBS decreasing the correlation length  with beam propagation  \cite{LushnikovRosePRL2004,LushnikovRosePlasmPhysContrFusion2006}. For low $Z$ plasma threshold for the collective FSBS is close to (\ref{I0convthresh}) \cite{LushnikovRosePRL2004}.
As $Z$ increases (which can be achieved by adding high $Z$ dopant), that threshold decreases below (\ref{I0convthresh}) and  might result in an increase of the BSBS threshold.

In conclusion, we identified a collective threshold for  BSBS instability of partially incoherent laser beam for ICF relevant plasma. Above that threshold the BSBS increment $\kappa_i$ is well
approximated by the sum of the collective like increment $\kappa_\sigma$
and RPA-like increment $\kappa_B$. We found that $\kappa_i$ is significantly below the BSBS  increment $\kappa_ {coherent}$ of the coherent laser beam.


We acknowledge helpful discussions with R. Berger and N. Meezan.
P.L and H.R. were supported by the New Mexico Consortium and Department of Energy
Award No. DE-SCOO02238.


\end{document}